# A New Type of Cipher: DICING_csb

Li An-Ping

Beijing 100085, P.R.China
apli0001@sina.com

**Abstract:** In this paper, we will propose a new type of cipher named DICING_csb, which is derived from our previous stream cipher DICING. It has applied a stream of subkey and an encryption form of block ciphers, so it may be viewed as a combinative of stream cipher and block cipher. Hence, the new type of cipher has fast rate like a stream cipher and need no MAC.
.

## 1. Introduction

The encryption form in a synchronous stream cipher usually is bitwise addition, namely, the ciphertext is made by bitwise adding (XOR) the plaintext with a binary sequence called keystream. A merit of this kind of ciphers is that there will be few of propagations for the errors in the communications. However, it is clear that in this additive encryption form the plaintext is easy to be falsified by other people. As a result, a synchronous stream cipher usually is equped a MAC (message authentication code) to protect the message from to be tampered.

In our algorithm DICING [1], the combining function had mainly applied keyed-Sboxes, which are often used in the block ciphers, we realize that it is possible to make a combinative of stream cipher and block cipher (CSB mode), so will be able to omit MAC in this way.

The main difference between the proposal cipher and the previous one DICING is the encryption forms. In the new one, the component $u_t$ of DICING will be applied as a role of a stream of subkeys, and the encryption means are mainly keyed-Sboxes, like one ordinary block cipher. The components used here are same in DICING, for the completeness, which will be repeated in this paper.

In the proposal cipher, we will apply a LFSR-*like* component called projector (Pr.). A projector consists of an element $\sigma_t$ called state from some finite field $GF(2^m)$ and an updating rule. The rule of updating states is that multiplying $\sigma_t$ with $x^k$, $k$ is an integer, namely,

$$\sigma_{t+1} = x^k \cdot \sigma_t. \tag{1.1}$$

The finite fields used in here are $GF(2^m)$, $m = 128$, or $127$. In other word, the operation *shift* in LFSR now is replaced by multiplying $x^k$ in the field $GF(2^m)$.

Like DICING, the key sizes in DICING_csb can be 128 bits or 256 bits, and the size of initial value may be taken as large as 256 bits, and the size of output of DICING is 128 bits.

In this paper the finite field $GF(2)$ is simply denoted as $\mathbb{F}$, and $\mathbb{F}[x]$ is the polynomial ring of unknown $x$ over the field $\mathbb{F}$. The symbols $\oplus$, $\otimes$ will represent the bitwise addition *XOR*, bitwise *and*, that is the operation $\&$ in $C$, and symbols $>>$, $<<$, $|$ and $\sim$ stand for the operations *right-shift, left-shift, concatenate* and *complement* respectively.

Suppose that $\zeta$ is a binary string, denoted by $\zeta[i]_{bit}$ and $\zeta[i,j]_{bit}$ the *i*-th bit and the segment from *i*-th bit to *j*-th bit respectively, and there are the similar expressions $\zeta[i]_{byte}$, $\zeta[i,j]_{byte}$ and $\zeta[i]_{word}$, $\zeta[i,j]_{word}$ measured in bytes and 32-bits words respectively, and if the meaning is explicit from the context, the low-index *bit*, *byte* and *word* will be omitted.

## 2. Construction

We will use two projectors $\Gamma_1$ and $\Gamma_2$, the first one acts a controller to control the updating of the second one, which will be used to form a stream of subkeys.

Denoted by $\alpha_t$ and $\omega_t$ the states of $\Gamma_1$ and $\Gamma_2$ in time $t$ respectively, which are based on the finite fields $\mathbb{E}_i$, $\mathbb{E}_i = \mathbb{F}[x]/p_i(x)$, $i = 1, 2$. $p_1(x)$ and $p_2(x)$ are the primitive polynomials with degree 127 and 128 respectively, which expression are given in the List 1. They satisfy the simple recurrence equations

$$\alpha_{i+1} = x^8 \cdot \alpha_i, \quad i = 0, 1, 2, \ldots. \tag{2.1}$$

The integer of the last eight bits of $\alpha_t$ is called the dice $D_t$, denoted by $d = 1 + (D_t >> 4)$, the state $\omega_t$ will be updated as

$$\omega_{t+1} = x^d \cdot \omega_t, \quad \text{for } t > 0. \tag{2.2}$$

Besides, we use a memorizes $u_t$ to assemble $\omega_t$,

$$u_t = u_{t-1} \oplus \omega_t, \quad \text{for } t > 0, \tag{2.3}$$

The initial values $\alpha_0$, $\omega_0$ and $u_0$ will be specified in the later.

Suppose that $\mathbb{K}$ is a finite field $GF(2^8)$, $\mathbb{K} = \mathbb{F}[x]/p(x)$, $p(x)$ is an irreducible polynomial of degree eight, which expression is given in the List 1. We define S-box $S_0(x)$ as

$$S_0(x) = 5 \cdot (x \oplus 3)^{127}, \quad x \in \mathbb{K}. \tag{2.4}$$

We also adopt the representation $S_0(\zeta)$ for a bytes string $\zeta$ to represent that S-box $S_0$ substitute each byte of the string $\zeta$.

The startup includes two subprocesses *keysetup* and *ivsetup*, where the basic materials as the secret key and key-size will be input and the internal states will be initialized. Besides, in the *keysetup* we will make two key-defined the S-boxes $S_1(x)$ and $S_2(x)$ from $S_0(x)$ and a diffusion transformation $L$. The process is as following.

For a string $\rho$ of 8 bytes, we define an 8-bits vector $V_\rho$ and a $8 \times 8$ matrix $M_\rho$:

$$V_\rho[i] = \rho[8i+i]_{bit}, 0 \leq i < 8, \quad M_\rho = T_u \cdot J \cdot T_l. \tag{2.5}$$

where $T_u = (a_{i,j})_{8\times 8}$ and $T_l = (b_{i,j})_{8\times 8}$ are the upper-triangular matrix and the lower-triangular matrix respectively,

$$a_{i,j} = \begin{cases} \rho[8i+j]_{bit} & \text{if } i < j, \\ 1 & \text{if } i = j, \\ 0 & \text{if } i > j, \end{cases} \qquad b_{i,j} = \begin{cases} \rho[8i+j]_{bit} & \text{if } i > j, \\ 1 & \text{if } i = j, \\ 0 & \text{if } i < j, \end{cases} \qquad (2.6)$$

and $J$ is a key-defined permutation matrix, for the simplicity, here take $J = 1$.

Suppose that $K$ is the secret key, let $K_c = K[0,23]_{byte} \oplus K[8,31]_{byte}$, if $|K| = 256$, else $K_c = K[0,15] | (K[0,7] \oplus K[8,15])$, $\lambda_i = K_c[(i-1)\times 8, 8i-1]$, $i = 1,2,3$. and define three affine transformations on $\mathbb{K}$

$$A(x) = M_{\lambda_1}(x), \quad B(x) = M_{\lambda_2}(x), \quad C(x) = M_{\lambda_3}(x), \qquad (2.7)$$

and a transformation $L$ on $\mathbb{K}^4$,

$$L = \begin{pmatrix} A & B & A & A \oplus B \\ B & A & A \oplus B & A \\ A & A \oplus B & A & B \\ A \oplus B & A & B & A \end{pmatrix}. \qquad (2.8)$$

Denoted by $v_i = \bigoplus_{0 \le k < 8} \lambda_i[k]_{byte}$, $i = 1,2,3$, and define two new S-boxes

$$S_1(x) = S_0(x \oplus v_1) \oplus v_2, \quad S_2(x) = C(S_0(x \oplus v_2) \oplus v_3), \quad x \in \mathbb{K}. \qquad (2.9)$$

Suppose that $\zeta$ is a string of $n$ bytes, if $n = 4k$ we also view it as a string of $k$ words, and write $L(\zeta)$ to represent that $L$ takes on the each word of $\zeta$. Simply, we denote

$$Q(\zeta) = L \cdot S(\zeta). \qquad (2.10)$$

In the *ivsetup*, the second step of the startup, the internal states will be initialized with the secret key and the initial value.

For a 32-bytes string $\zeta$ we define a bytes permutation $\phi$: $\zeta^\phi = \phi(\zeta)$, $\zeta^\phi[i] = \zeta[4i \mod 31]$, for $0 \le i < 31$, and $\zeta^\phi[31] = \zeta[31]$. Let $\tilde{K} = K$ if $|K| = 256$ else $\tilde{K} = K | (\sim K)$, denoted by $\tilde{K}_0 = \tilde{K}$, $\tilde{K}_i = \tilde{K}[8i, 31]_{byte} | \tilde{K}[0, 8i-1]_{byte}$, $i = 1,2,3$. We define the functions recurrently

$$F(\zeta) = Q(\phi(\zeta)), \quad F_0(\zeta) = F(\zeta) \oplus \tilde{K}_0, \quad F_i(\zeta) = F(F_{i-1}(\zeta)) \oplus \tilde{K}_i. \quad i = 1,2,3. \qquad (2.11)$$

Suppose that $IV$ is the initial value of 32-bytes, $e$ is the base of natural logarithm and $c$ the integral part of $e \cdot 57!$, and $\xi_i$, $0 \le i \le 3$, are three 32-bytes strings defined as

$$\xi_0 = F_3(IV \oplus c), \quad \xi_i = F_3(\xi_{i-1} \oplus c), \quad i = 1, 2. \tag{2.12}$$

Let $\eta = \xi_0[0,15] \oplus \xi_0[16,31]$, which will be employed in the encryption, and the internal states are initialized respectively as following

$$u_0 = \xi_1[0,15], \quad \alpha_0 = \xi_1[128, 254]_{bit}, \quad \omega_0 = \xi_2[0,15], \tag{2.13}$$

If $\xi_2[0,15] = 0$, the states $\omega_0$ will be re-set as

$$\omega_0 = \xi_2[16, 31]. \tag{2.14}$$

*Note.* For a secret key, there is at most one $IV$ such that $\xi_2 = 0$.

In the proposal cipher DICING_csb, the sequence $\{u_t\}$ will play a flow of subkeys. After initializing, the process enters the recurrence part of encryption/decryption, in which including the sub-process of updating the states, that is, making the stream of subkeys $\{u_t\}$. Denoted by $\{x_t\}_{t>0}$ and $\{y_t\}_{t>0}$ the sequences of plaintext and ciphertext respectively, the encryption function is defined as

$$y_t = S_2(Q(x_t \oplus u_t) \oplus Q(\eta)) \oplus u_t. \tag{2.15}$$

We have summarized the whole process in a sketch as Fig. 1.

The Sketch of Encryption Process

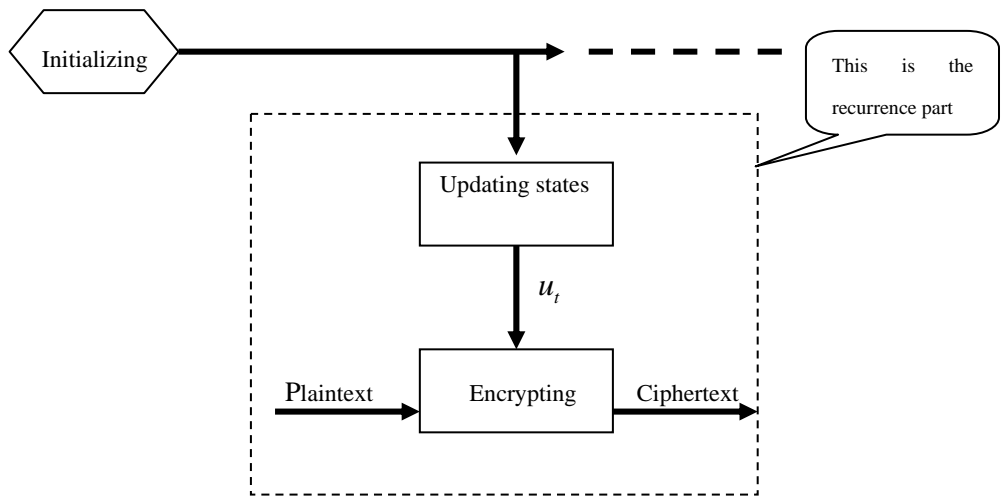

Fig.1

List of the Primitive Polynomials used

| Polynomials | Expression |
|---|---|
| $p(x)$ | $x^8 + x^6 + x^5 + x + 1$ |
| $p_1(x)$ | $x^{127} + (x^{89} + x^{41} + 1)(x^3 + 1)$ |
| $p_2(x)$ | $x^{128} + (x^{96} + x^{67} + x^{32} + 1)(x^3 + 1)$ |

List 1

## 3. Security Analysis

The analysis for DICING_csb as a stream cipher will be similar to the one for DICING, refer to see the paper [1]. Besides, as a block cipher, the encryption form of DICING_csb is not usual iterative one, so the traditional analyses for the block ciphers of iterative mode will not be feasible. On the other hand, not as additive stream ciphers are vulnerable for plaintext-recovery attacks such that a IV may be applied only one time, DICING_csb may use a IV as most as $2^{16}$ times. This also means that a keystream $u_t$ may be employed several times in the CSB encryption form, which will refer to an alternative version, see the section 5. If intend to apply a IV more than $2^{16}$ times, then in encryption function should be added a more round in order to enlarge the range of diffusion, as a cost, the encrypting rate will be added more about 2 cycles/byte.

It maybe should be mentioned that we have reduced two Pr.'s from DICING for we think that in this encryption form the requirement for the period of the sequence $\{u_t\}$ may be relaxed, here the period of $\{u_t\}$ is no less than $(17 \cdot 2^{126} - 1)(2^{128} - 1)$.

## 4. Implementation

In the platform of 32-bit Windows OS and AMD Athlon(tm) 64 x2 Dual Core processor 3600+, 2.00G Borland C++ 5.0, the performance of DICING_csb is as following

Report of Performance

| Encryption | | Decryption | |
|---|---|---|---|
| Sub-processes | Time | Sub-processes | Time |
| Keysetup | 8340 *cycles* | Keysetup | 12100 *cycles* |
| IVsetup | 4280 *cycles* | IVsetup | 4340 *cycles* |
| Encryption rate | 8.4 *cycles/byte* | Decryption | 8.4 *cycles/byte* |

## 5. Some variants of DICING_csb

I.  There is an alternate updating rule for the states $\alpha_t$ and $\omega_t$ as following

$$\alpha_{t+1} = x^{16} \cdot \alpha_t, \quad \text{for } t > 0. \tag{5.1}$$

Denoted by $d_i = 1 + (\alpha_{t+1}[i]_{byte} \,\&\, 15)$, $0 \leq i < 16$, the states $\omega_t$ are updated as

$$\omega_{16t+i+1} = x^{d_i} \cdot \omega_{16t+i}, \quad 0 \leq i < 16, \quad \text{for } t \geq 0. \tag{5.2}$$

With the updating rule above, the encrypting/decrypting rate will be fasted to $6.8\, cycles/byte$ in the case of larger size of massage. We call the rules (5.1) and (5.2) as *lotting*.

II. As mentioned in DICING [1], we may substitute a Pr. $\hat{\Gamma}$ of a finite field $GF(2^{256})$ for the two Pr.'s $\Gamma_1$ and $\Gamma_2$. Suppose that $\zeta_t$ is the state of $\hat{\Gamma}$ in the time $t$, which is updated as following

$$\zeta_{t+1} = x^{r_t} \cdot \zeta_t, \quad \text{for } t > 0, \tag{5.3}$$

where $r_t = 1 + \zeta_t[252, 255]_{bit}$.

Let $\{w_t\}$ be a sequence of 32-bytes words, which is defined recurrently $w_{t+1} = w_t \oplus \zeta_{t+1}$, denoted by $w'_t = w_t[0,15]$, $w''_t = w_t[16,31]$, the encryption function is defined as

$$y_t = S_2(Q(x_t \oplus w''_t) \oplus Q(\eta)) \oplus w'_t. \tag{5.4}$$

III. In the section 3, we mentioned that for DICING_csb a IV may be permitted to use several times, and a keystream $u_t$ (or $w_t$) also may be iteratively employed in encryption a number of times, say, 16 times, then the encryption function defined in (2.15) will be changed into that

$$y_{16t+i} = S_2(Q(x_{16t+i} \oplus u_t) \oplus Q(\eta)) \oplus u_t, \quad t \geq 0, \quad 0 \leq i < 16. \tag{5.5}$$

Clearly, in this way, with increase the usage of a keystream $u_t$ (or $w_t$), the encryption/decryption rate will be much improved, and catch up with a stream cipher, the form more like a block cipher.

## 6. Conclusion

The proposal cipher can be viewed as a combinative of stream cipher and block cipher. It assimilates the good qualities of stream ciphers in the speed and block ciphers in the secure. It is able to serve as a synchronous stream cipher or a block cipher, and there will no need to equip a MAC when it is applied as a synchronous stream cipher. While it is applied as block cipher, it will still require a IV to initialize the internal states, however, this requirement is easy to be simply satisfied, for example, the name or the date of files may be taken as the IV'values.